\documentclass{eas}
\usepackage{graphicx}
\begin{document}
\runningtitle{A. Sozzetti: Astrometry and Exoplanets: the Gaia Era, and Beyond}
\title{Astrometry and Exoplanets: the Gaia Era, and Beyond} 
\author{A. Sozzetti}\address{INAF - Osservatorio Astronomico di Torino, 
Strada Osservatorio, 20 - 10025 Pino Torinese Italy}
%
%
\begin{abstract}

The wealth of information in the Gaia catalogue of exoplanets will constitute a 
fundamental contribution to several hot topics of the astrophysics of planetary systems. 
I briefly review the potential impact of Gaia micro-arsec astrometry 
in several areas of exoplanet science, discuss what key follow-up observations might 
be required as a complement to Gaia data, and shed some light on the role of next 
generation astrometric facilities in the arena of planetary systems. 

\end{abstract}
\maketitle
\section{Introduction}

The study of exoplanets orbiting solar-type stars has emerged in the last decade to be one 
of the most exciting new areas of astronomy and planetary science. Since 1995, almost 500 
planets outside the solar system have been discovered, mostly by the radial-velocity method 
(a technique that can only put lower limits on the mass of any detected companion). 
Collectively, they span a huge range of masses and orbital distances, are commonly found in 
multiple systems (the record-holder being HD 10180 with its 7-planet system), have been detected 
around star of varied spectral types and evolutionary stages, and have led to major revisions in our ideas of how and where 
planets form and what their structure is. The majority of these planets are gas giants, 
but more than 70 are of Neptune mass and lower. Ninety-seven (four of which in multiple-planet systems) 
are known to transit across the disk of their primary star, and it is for these that we 
have physical parameters such as radius and mass. A subset of the close-in transiting giants 
has now been detected directly at secondary eclipse and at various orbital phases by the 
Spitzer Infrared Space Telescope and the Hubble Space Telescope, providing the first 
measurements of extrasolar planetary atmospheres and compositions. To interpret these data, 
theorists have developed models for planet formation, orbital interaction and dynamics, 
evaporation due to stellar irradiation, atmospheric circulation and global heat transport, 
atmospheric structure and spectra, phase light curves, the equations of state of their 
interiors, molecular chemistry, radius evolution, and tidal effects, to name only a few topics. 

Their emerging properties in the near future may be put on much more solid statistical grounds, 
thanks to an going and soon to become larger flow of observational data collected with a 
variety of techniques which help answer some of the most outstanding questions what some refer 
to now as the science of ``comparative exoplanetology''. Among the vast array of techniques for 
planet detection and characterization, the potential of high-precision astrometric measurements 
is still mostly unexplored. After decades of `blunders' (e.g., Sozzetti 2010, and references therein), 
micro-arcsecond ($\mu$as) astrometry is now coming of age, which will allow this 
technique to obtain in perspective the same successes of the Doppler method, for which 
the improvement from the km s$^{-1}$ to the m s$^{-1}$ precision opened the doors for 
ground-breaking results in exoplanetary science. 
I briefly review here what promises the next decade holds for astrometry and exoplanets. 

\section{Gaia and extrasolar planets}

Gaia's ambitious science case (e.g., Perryman et al. 2001), 
wishing to address breakthrough problems in Milky Way astronomy, 
great impact is the astrophysics of planetary systems (e.g., Casertano et al. 2008), 
in particular when seen as a complement to other techniques for planet detection 
and characterization (e.g., Sozzetti 2010). 
Given the characteristics of Gaia's magnitude--limited survey 
(see e.g. the reviews by Prusti and Charvet, this volume), and the non-competitiveness of 
its spectroscopic and photometric capabilities with facilities devoted to 
high-precision radial-velocity measurements (e.g., Pepe \& Lovis 2008) 
or transit photometry (e.g., Sozzetti et al. 2010), the mission's 
potential contribution to exoplanets science must be purely gauged in terms of its
astrometric capabilities. 

\subsection{The Gaia Double-Blind Tests Campaign}

\begin{figure}
\centering
\includegraphics[width=0.65\textwidth]{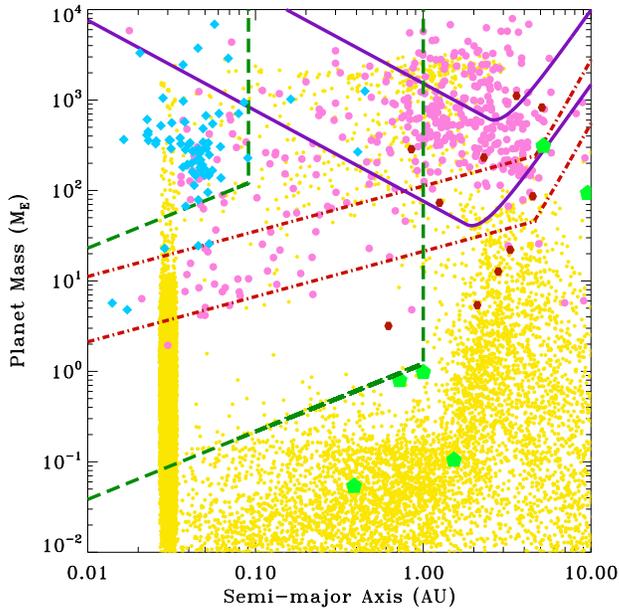}
\caption{Gaia exoplanets discovery space (blue curves) compared to that of Doppler (red lines) 
and transit (green curves) techniques. 
Detectability curves are defined on the basis of a 3-$\sigma$ criterion for signal detection 
(see Sozzetti 2010 for details).
The upper and lower blue solid curves are for Gaia astrometry with $\sigma_\mathrm{A} = 10$ $\mu$as,
assuming a 1-$M_\odot$ primary at 200 pc and a 0.4-$M_\odot$ M dwarf at 25 pc, respectively, 
and survey duration set to 5 yr. 
The pink filled circles indicate the inventory of Doppler-detected exoplanets as of May 2010. Transiting systems
are shown as light-blue filled diamonds. Red hexagons are planets detected by microlensing.
Solar System planets are also shown (large green pentagons). The small yellow dots represent a theoretical
distribution of masses and final orbital semi-major axes (Ida \& Lin 2008).}

\label{Gaiafig}
\end{figure}

In the large-scale, double-blind test campaign carried out to estimate the potential of
Gaia for detecting and measuring planetary systems, Casertano et al. (2008) showed that, 
considering bright stars ($V < 13$) uniformly observable with the best-achievable single-measurement 
astrometric precision ($\sigma_\psi\simeq 10-15$ $\mu$as), 
Gaia could discover and measure massive giant planets ($M_{\mathrm{p}} \geq $2--3 $M_\mathrm{J}$)
with $1<a<4$ AU orbiting solar-type stars as far as the nearest star-forming regions,
as well as explore the domain of Saturn-mass planets with similar
orbital semi-major axes around late-type stars within 30--40~pc (see Figure~\ref{Gaiafig}). These results can be
used to infer the number of planets of given mass and orbital separation that can be detected and measured by Gaia,
using Galaxy models and the current knowledge of exoplanet frequencies. By inspection of the two sub-tables in
Table~\ref{nsyst}, one then finds that Gaia's main strength will be its ability to accurately measure
orbits and masses for thousands of giant planets, and to perform coplanarity measurements for a few
hundred multiple systems with favorable configurations. 

Finally, a word of caution is in order on the face of the
possible degradations in the expected Gaia astrometric precision
on bright stars ($V < 13$). As the refinement of the overall Gaia error budget progresses, 
including a better understanding of some effects which can potentially affect 
centroiding as well as systematic uncertainties to a significant degree, such as the 
charge transfer inefficiency effect due to solar radiation bombardment of Gaia CCDs at L2 
(see e.g. Prod$^\prime$homme and Pasquier, this volume), a worsening of Gaia astrometric 
performance might impact more or less seriously the Gaia science case for exoplanets. Casertano et al. (2008) 
have shown how the number of useful stars, and of detectable and measurable single- and 
multiple-planet systems would decrease as $\sigma_\psi$ increases 
(assuming that the number of objects scales with the cube of the radius, in pc, 
of a sphere centered around the Sun). 
In the context of that exercise, a factor 2 degradation in astrometric precision would 
impact most of Gaia exoplanet science case. Until the true Gaia performance on real data will be known, 
it will be necessary to closely follow further developments 
in the understanding of the technical specifications of Gaia and its instruments, and of its
observation and data analysis process, in order to revisit
these issues as needed in the future. 

\begin{table}[tbh]
\begin{minipage}{0.5\linewidth}
\begin{center}
   \renewcommand{\arraystretch}{1.4}
   \setlength\tabcolsep{7pt}
{\tiny
      \begin{tabular}{|c|c|c|c|c|c|c|}
       \hline\noalign{\smallskip}
       $\Delta d$ & $N_\star$  & $\Delta a$ & $\Delta M_p$ &  $N_{\rm d}$ &  $N_{\rm m}$ \\
       (pc) & & (AU) & ($M_J$) & & \\
              \noalign{\smallskip}
       \hline
       \noalign{\smallskip}
0-50 & $1\times10^4$ & 1.0 - 4.0 & 1.0 - 13.0 & $1400$ & $ 700$\\ \hline 
50-100 & $5\times10^4$ & 1.0 - 4.0 & 1.5 - 13.0 & $2500$ & $ 1750$\\ \hline 
100-150 & $1\times10^5$ & 1.5 - 3.8 & 2.0 - 13.0& $2600$ & $ 1300$\\\hline 
150-200 & $3\times10^5$ & 1.4 - 3.4 & 3.0 - 13.0& $2150$ & $ 1050$\\\hline 

   \end{tabular}
}
\end{center}
\centering\large (a)
\end{minipage}
\hspace{0.5cm}
\begin{minipage}{0.5\linewidth}
\begin{center}
   \renewcommand{\arraystretch}{1.4}
   \setlength\tabcolsep{7pt}
{\tiny
      \begin{tabular}{|l|c|}
       \hline\noalign{\smallskip}
        Case & N. of systems \\
              \noalign{\smallskip}
       \hline
       \noalign{\smallskip}
Detection & $\sim 1000$\\ \hline
Orbits and masses to  & \\ 
$<15-20\%$ accuracy & $\sim 400-500$ \\ \hline
Successful  & \\
coplanarity tests & $\sim 150$\\ \hline
   \end{tabular}
}
\end{center}
\centering\large (b)
\end{minipage}
\caption{Left: Number of giant planets of given ranges of mass ($\Delta M_p$) and orbital separation 
($\Delta a$) that could be detected ($N_d$) and measured ($N_m$) by Gaia, 
as a function of increasing distance ($\Delta d$) and stellar sample ($N_\star$). 
Right: Number of planetary systems that Gaia could potentially detect, measure,
and for which coplanarity tests could be carried out successfully. See Casertano et al. (2008) 
for details.}
\label{nsyst}
\end{table}

\subsection{Astrometric Modeling of Planetary Systems} 

The problem of the correct determination of the astrometric orbits of planetary systems 
using Gaia data (highly non-linear orbital fitting procedures, large numbers 
of model parameters) will present many difficulties. For example, it will be necessary
to assess the relative robustness and reliability of different procedures for orbital fits, 
with a detailed understanding of the statistical properties of the uncertainties
associated with the model parameters. For multiple systems, a trade-off will have to be 
found between accuracy in the determination of the mutual inclination angles between
pairs of planetary orbits, single-measurement precision and redundancy in the number
of observations with respect to the number of estimated model parameters. It will 
be challenging to correctly identify signals with amplitude close to the measurement
uncertainties, particularly in the presence of larger signals induced by other companions 
and/or sources of astrophysical noise of comparable magnitude. Finally, for 
systems where dynamical interactions are important (a situation experienced 
already by Doppler surveys), fully dynamical fits involving
an n-body code might have to be used to properly model the Gaia astrometric data and
to ensure the dynamical stability of the solution (see Sozzetti 2005).
All the above issues could significantly impact Gaia's planet detection 
and characterization capabilities. For these reasons, a Development Unit (DU), 
within the pipeline of Coordination Unit 4 (object processing) of the Gaia Data 
Processing and Analysis Consortium, has been
specifically devoted to the modelling of the astrometric signals produced by planetary
systems. The DU is composed of several tasks, which implement multiple robust procedures 
for (single and multiple) astrometric orbit fitting (such as Markov Chain Monte
Carlo algorithms) and the determination of the degree of dynamical stability
of multi-planet systems. 

\subsection{The Gaia Legacy}

Gaia's main contribution to exoplanet science will be its unbiased census of planetary systems 
orbiting hundreds of thousands nearby ($d< 200$ pc), relatively bright ($V \leq 13$) stars 
across all spectral types, screened with constant astrometric sensitivity. As a result, 
the actual impact of Gaia measurements in exoplanets science is broad, and rather structured. 
The Gaia data have the potential to: a) significantly refine our understanding of the statistical properties
of extrasolar planets; 
b) help crucially test theoretical models of gas giant planet formation and migration; 
c) achieve key improvements in our comprehension of important aspects of
the formation and dynamical evolution of multiple-planet systems; 
d) aid in the understanding of direct detections of giant extrasolar planets; 
e) provide important supplementary data for the optimization of the 
target selection for future observatories aiming at the direct detection and spectral characterization of 
habitable terrestrial planets. 
Finally, ongoing studies (Sozzetti et al., in preparation) 
are now focusing on the detailed understanding of the 
planet discovery potential of Gaia as far as low-mass and post-main-sequence 
stars and are concerned. 

\subsection{Synergies Between Gaia and Other Programs}

The broad range of applications to exoplanets science is such that Gaia data can be 
seen as an ideal complement to (and synergy with) many ongoing and future observing programs devoted 
to the indirect and direct detection and characterization of planetary systems, both from the ground and in space. 
Gaia will contribute critically, for example, to the definition of 
input catalogues for proposed quasi-all-sky photometric transit surveys (PLATO); 
it will inform direct imaging surveys (e.g., SPHERE/VLT, EPICS/E-ELT) and spectroscopic characterization projects 
about the epoch and location of maximum brightness of (primarily non transiting) exoplanets, 
in order to estimate their optimal visibility, and will help in the modeling and interpretation 
of giant planets' phase functions and light curves. 

Another critical aspect will concern the large effort in terms of ground-based follow-up activities 
to improve the characterization of astrometrically detected systems (and possibly those 
found transiting). For example, high-precision radial-velocity campaigns (both at visible and 
infrared wavelengths) will be a necessary complement, with the three-folded aim of improving the phase sampling of the 
astrometric orbits found by Gaia, extending the time baseline of the observations (to 
put stringent constraints on or actually characterize long-period companions), and 
search for additional, low-mass and/or short-period components which might have been missed by 
Gaia due to lack of sensitivity. 

\section{High-Precision Astrometry for Exoplanets: Beyond Gaia}

In addition to Gaia, there are several astrometric projects devoted to exoplanet
detection and characterization (e.g, Sozzetti 2010; Malbet et al. 2010) both 
from the ground and in space, which are either ongoing, about to start, or under development/study.  
They vary in accuracy level and number of potential targets. 

Ground-based relative astrometry programs with large telescopes and adaptive optics (VLT/FORS2, 
CAPSCam, STEPSS) are focusing on relatively small numbers (a few tens to one hundred) of very 
faint targets (late M, L, and T dwarfs) in dense stellar fields. The expected long-term 
precision of these observatories will not exceeed $0.1-1$ mas. The performance of instrumentation for 
coronagraphic astrometry has been studied (e.g., Digby et al. 2006) and found not to be competitive 
in term of achievable precision, for the time being. Ground-based dual-star 
interferometry projects (VLTI/PRIMA, ASTRA) are designed to perform narrow-angle
interferometric astrometry of a very bright target and one moderately faint 
reference star separated by up to 1$^\prime$ 
with expected accuracies better than 100 $\mu$as. Also in this case, the target lists 
will be composed by typically 100 stars. 

In space, HST/FGS can still guarantee an astrometric precision of $0.3-0.5$ mas (exceeding that of 
Hipparcos), for the improved characterization of selected, 
suitable systems known to host massive planets from Doppler observations. The long-term ($\sim20$ years) 
SIM/SIM-Lite optical interferometry project, capable of narrow-field astrometry with 1-$\mu$as precision, 
has recently not been recommended for funding by NASA's Decadal Survey. No space-borne astrometric 
facility is now foreseen to exceed Gaia's astrometric performance during or after the end of Gaia 
operational phase. Studies of the projected performance of high-precision narrow-angle 
coronagraphic astrometry in space are still at the infancy status.

\section{Conclusions}

The Gaia mission is now set to establish the European leadership in high-precision astrometry for the next 
decade. The largest compilation of high-accuracy astrometric orbits of giant 
planets, unbiased across all spectral types up to $d\simeq200$ pc, 
will allow Gaia to crucially contribute to several aspects of planetary systems astrophysics 
(formation theories, dynamical evolution), in combination with present-day and future 
extrasolar planet search programs. 




...

\begin{thebibliography}{}%
\bibitem[Casertano et al. (2008)]{casertano08}
Casertano, S., Lattanzi, M. G., Sozzetti, A., et al. 2008, A\&A, 482, 699
\bibitem[Digby et al. (2006)]{digby06}
Digby, A. P., Hinkley, S., Oppenheimer, B. R., et al. 2006, ApJ, 650, 484
\bibitem[Ida \& Lin (2008)]{ida08}
Ida, S., \& Lin, D. N. C. 2008, ApJ, 673, 487
\bibitem[Malbet et al. (2010)]{malbet10} 
Malbet, F., Sozzetti, A., Lazorenko, P. et al. 2010, ASP Conf. Ser., 430, 84
\bibitem[Pepe \& Lovis (2008)]{pepe08}
Pepe, F., \& Lovis, C. 2008, Physica Scripta, 130, 014007
\bibitem[Perryman et al. (2001)]{perryman01}
Perryman, M. A. C., et al. 2001, A\&A, 369, 339
\bibitem[Sozzetti (2005)]{sozz05}
Sozzetti, A. 2005, PASP, 117, 1021
\bibitem[Sozzetti (2010)]{sozz10}
Sozzetti, A. 2010, EAS Publication Series, 42, 55
\bibitem[Sozzetti et al. (2010)]{sozzetal10}
Sozzetti, A., Afonso, C., Alonso, R., et al. 2010, ASP Conf. Ser., 430, 45
\end{thebibliography}
\end{document}